\documentclass[proof]{pasj00}
\SetRunningHead{Jihye Kang}{Force-freeness of a solar emerging magnetic field}

\usepackage{times}
\usepackage{epsfig}

\begin{document}

%%\title{Force-Freeness of Emerging Magnetic Field on the Sun Derived from Virial Theorem}
\title{INVESTIGATION OF FORCE-FREENESS OF A SOLAR EMERGING MAGNETIC FIELD VIA APPLICATION OF THE VIRIAL THEOREM TO MHD SIMULATIONS}

\author{Jihye \textsc{Kang}}
\affil{School of Space Research, Kyung Hee University,
1, Seocheon-dong, Yongin, 446-701, Korea}
\email{siriustar@khu.ac.kr}

\and

\author{Tetsuya \textsc{Magara}}
\affil{Department of Astronomy and Space Science, School of Space Research, Kyung Hee University,
1, Seocheon-dong, Yongin, 446-701, Republic of Korea}


\maketitle

\begin{abstract}
Force-freeness of a solar magnetic field is a key to reconstructing invisible coronal magnetic structure of an emerging flux region on the Sun where active phenomena such as flares and coronal mass ejections frequently occur. We have performed magnetohydrodynamic (MHD) simulations which are adjusted to investigate force-freeness of an emerging magnetic field by using the virial theorem. Our focus is on how the force-free range of an emerging flux region develops and how it depends on the twist of a pre-emerged magnetic field. As an emerging flux region evolves, the upper limit of the force-free range continuously increases while the lower limit is asymptotically reduced to the order of a photospheric pressure scale height above the solar surface. As the twist becomes small the lower limit increases and then seems to be saturated. We also discuss the applicability of the virial theorem to an evolving magnetic structure on the Sun. {\bf A manuscript with high-resolution figures is found at http://web.khu.ac.kr/$\sim$magara/index.html. Here `$\sim$' is a tilde.} 
\end{abstract}

\section{Introduction}

Active phenomena such as flares and coronal mass ejections (CMEs) are frequently observed on the Sun and it is now believed that they are caused by the release of free magnetic energy in the solar corona. An emerging flux region with intense magnetic flux in it is a typical area where active phenomena occurs, which is called an {\it active region} (Forbes 2000; Shibata \& Magara 2011; Chen 2011). In order to understand the physical mechanism for producing active phenomena it is important to know the coronal magnetic structure of an emerging flux region, although it is still difficult to derive a detailed configuration of a coronal magnetic field only by observations. We therefore need to reconstruct a coronal magnetic structure using observable quantities. A photospheric magnetic field is one of these observable quantities and the vector information on a photospheric field has widely been used to reconstruct a magnetic structure with free magnetic energy in the corona (Wang et al. 1994; Cameron \& Sammis 1999; R{\'e}gnier \& Amari 2004; Wang 2006).

One of the ways to reconstruct a coronal magnetic structure is to solve the so-called force-free equation given by (e.g. Sturrock 1994)
\begin{equation}
\nabla\times {\bf B} = \alpha {\bf B}.
\label{eq:ff}
\end{equation}
Here $\alpha$ is a function of position (force-free parameter). The solutions of Equation (1) give potential fields (PFs) when $\alpha \equiv 0$, while they give linear force-free fields (LFFFs) when $\alpha$ is a non-zero constant. In the most general case where $\alpha$ changes spatially the solutions give nonlinear force-free fields (NLFFFs) (Woltjer 1958; Molodenskii 1969; Aly 1984; Berger 1985). Among these three kinds of force-free fields, PFs can give the minimum energy of a coronal magnetic structure. On the other hand, both LFFFs and NLFFFs contain free magnetic energy in the corona although LFFFs are obtained under a more restricted condition on $\alpha$ than NLFFFs, suggesting that LFFFs have difficulty in reconstructing a coronal magnetic structure full of variety formed on the Sun.

Reconstruction of a coronal magnetic structure using NLFFFs has well been studied, providing various practical methods such as the optimization method (Wheatland, Sturrock \& Roumeliotis 2000; Wiegelmann 2004), magnetofrictional method (Yang, Sturrock \& Antiochos 1986; McClymont \& Mikic 1994; Roumeliotis 1996; McClymont, Jiao \& Mikic 1997; Inoue et al. 2011), Grad-Rubin method (Grad \& Rubin 1958; Amari et al. 1997; Amari, Boulmezaoud \& Mikic 1999) and Green's function method (Yan \& Sakurai 1997; 2000). A recent good review on these methods is found in Wiegelmann \& Sakurai (2012).

%%When we use one of those force-free reconstruction methods, we should recall two assumptions for a force-free approximation. One is that the gas pressure is negligible compared to the magnetic pressure (plasma $\beta$ is much smaller than 1). Another is that plasma motions do not affect a magnetic field significantly, which guarantees that a magnetic field is close to a static state.
%%are more or less deviated from static states including force-free states. 

An emerging flux region that produces active phenomena is initially formed through the emergence of intense magnetic flux below the solar surface, called {\it flux emergence}. One of the results from a flux-emergence magnetohydrodynamic (MHD) simulation suggests that the coronal magnetic structure of an emerging flux region is divided into inner quasi-static part and outer expanding (dynamic) part (Magara \& Longcope 2003). A question then arises: how well a coronal magnetic structure formed via flux emergence is reconstructed using a force-free field; in other words how the force-free range of an emerging flux region develops.

%%
%% and after that they continue to evolve. In the present study we reproduce the emergence and the succeeding evolution of a magnetic field via an magnetohydrodynamic (MHD) simulation and investigate how much a magnetic structure dynamically formed on the Sun is close to a force-free state in a quantitative manner.

When a magnetic field is force-free, the magnetic energy stored in a semi-infinite region above the solar surface is given by the following surface integral (e.g. Priest 1982):
\begin{equation}
E_{vr}={1 \over 4 \pi} \int_{z=z_0}(xB_x+yB_y)B_z dxdy,
\label{eq:ev}
\end{equation}
where $x$ and $y$ form a horizontal plane while $z$ is directed upward ($z_0=0$ corresponds to the solar surface, i.e. the photosphere). Equation (2) has widely been used to estimate the magnetic energy of a coronal structure from a photospheric magnetic field (Gary et al. 1987; Sakurai 1987; Klimchuk et al. 1992; Metcalf et al. 1995; McClymont et al. 1997; Wheatland \& Metcalf 2006; R{\'e}gnier \& Priest 2007). In the present study we have performed a series of flux-emergence MHD simulations to reproduce several emerging flux regions with different magnetic configurations. We then calculate the magnetic energy of a coronal structure directly via the following volume integral:
\begin{equation}
E_{m}=\int_{z \ge z_0} {B^2 \over 8\pi} dx dy dz.
\label{eq:em}
\end{equation}
If a magnetic structure formed via flux emergence is in a force-free state, $E_{vr}$ and $E_{m}$ should be matched. In this respect, we have to mention the ambiguity of Equation (2), that is, the value of $E_{vr}$ generally depends on the location of the origin of a Cartesian coordinate system when a magnetic field is not force-free (Wheatland \& Metcalf 2006). Since a dynamically evolving magnetic structure reproduced by a flux-emergence simulation is not in a force-free state, it seems that Equation (2) cannot be applied to a flux-emergence simulation. To avoid this problem we have adjusted our simulations to remove the ambiguity of Equation (2), which is explained in detail in section 2. Accordingly we can use $E_{vr}$ as a tool to investigate force-freeness of an emerging flux region.

The organization of this paper is as follows: In Section 2 we describe the setup of the simulations focused on force-freeness of an emerging flux region and explain how to avoid the ambiguity of Equation (2). Section 3 presents comparisons between $E_{vr}$ and $E_{m}$ of several emerging flux regions reproduced by the simulations. In Section 4 we discuss how the force-free range of an emerging flux region evolves and how it depends on the twist of a pre-emerged magnetic field. We also discuss the applicability of the virial theorem to an evolving magnetic structure formed via flux emergence on the Sun.

\section{Setup of MHD simulations focused on force-freeness}

We have performed a series of three-dimensional MHD simulations in a Cartesian coordinate system to reproduce several emerging flux regions with different magnetic configurations. The simulations have been done by solving ideal MHD equations given by
\begin{equation}
{\partial \rho \over \partial \mit t}
 + \bf{\nabla} \cdot \left({\rho
 \bf v}\right) = \rm 0,
 \label{eq:cont}
\end{equation}
\begin{equation}
\rho \left[{\partial \bf v \over \partial \mit t}
 + \left({\bf v \cdot \bf \nabla}\right) \bf v \right]
 = -\nabla \mit P
 + {\rm 1\over 4\pi}
 \left({\nabla\times\bf B}\right) \times \bf B - \rho {\mit
g_0} \hat{\bf z},
\end{equation}
\begin{equation}
{\partial \bf B \over\partial \mit t}
 = \nabla \times \left({\rm \bf v \times \bf B}\right),
\end{equation}
\begin{equation}
{\frac{\partial \,P}{\partial \,t}}+\nabla \cdot
\left({P\bf v}\right)=-\left({\gamma -1}\right)P\nabla
\cdot \bf v,
\end{equation}
and
\begin{equation}
P={{\rho \Re \mit T}\over \mu},
\end{equation}
where $\rho$, $\bf{v}$, $\bf{B}$, $P$, $g_0$, $\gamma$, $\mu$, $\Re$,
and $T$ represent the gas density, fluid velocity, magnetic field, gas pressure, gravitational acceleration, adiabatic index ($\gamma=5/3$ is assumed), mean molecular weight ($\mu=0.6$ is assumed), gas constant and temperature, respectively. Table 1 shows the units of physical quantities used in the present study. The simulations have been performed within a domain $(-200, -200, -10) < (x, y, z) < (200, 200, 190)$ where the $z$-axis is directed upward and $z=0$ corresponds to the solar surface.  We adopt a non-uniform grid system $(\Delta x, \Delta y, \Delta z) = (0.1, 0.2, 0.1)$ for $(-8,-12,-10) < (x, y,z) < (8,12,15)$, while $\Delta x, \Delta y$ and $\Delta z$ increase toward $4$ as $|x|, |y|$ and $z$ increase.

As an initial state, we place an isolated magnetic flux tube horizontally below the surface, whose axis is aligned with the $y$-axis, crossing the $z$-axis at $z=-4$. This flux tube is in mechanical equilibrium with a background atmosphere stratified under uniform gravity, which is the same as that in our previous works (An \& Magara 2013; Lee \& Magara 2014). The magnetic field composing a flux tube is characterized by a Gold-Hoyle profile, given by
\begin{equation}
{\bf B}={B}_{0}{\frac{-b\ r\ \hat{\theta }+\hat{y}}{1+{b}^{2}\
{r}^{2}}},
\end{equation}
where $\hat{y }$ and $\hat{\theta}$ denote the axial and azimuthal directions of a flux tube, while $r$, $B_0$ and $b$ represent the radial distance from the axis, the strength of a magnetic field at the axis and the twist parameter, respectively.

In the present study we investigate the evolution of five different flux tubes with the common background atmosphere (the photospheric pressure scale height is given by $H_{ph}=P_{ph}/(\rho_{ph} g_0)$). These flux tubes have the same radius ($r=2$) and the same magnetic pressure at the interface with the background atmosphere (the gas pressure inside each flux tube is reduced to maintain pressure equilibrium at the interface) while they have different values of $b$ and $B_0$: $b=0.2$ (Extremely Weak Twist, EWT case), $b=0.35$ (Weak Twist, WT case), $b=0.5$ (Medium Twist, MT case), $b=0.8$ (Strong Twist, ST case) and $b=1.0$ (Extremely Strong Twist, EST case). More detailed information on these flux tubes is presented in Table 2. Figure 1 shows the distributions of gas pressure, magnetic pressure, gas density and temperature along the $z$-axis in all the cases.

To drive a simulation, we impose the following velocity perturbation to a flux tube during $0<t<t_r$ in each case:
\begin{equation}
{v}_{z} = \left\{{\begin{array}{cl}
\frac{{v}_{0}}{2}\cos\left({2\pi \frac{y}{\lambda }}\right)\sin
\left({\frac{\pi }{2} \frac{t}{{t}_{r}}}\right) & \quad \mbox{for $
\left|{y}\right| \le \frac{\lambda }{2}$} \\
\frac{{v}_{0}}{2}\cos\left(2\pi \frac{y-\left[{2L-\frac{\lambda }{2}}
\right] \frac{\left|{y}\right|}{y}}{4L-2\lambda }\right)\sin
\left({\frac{\pi }{2} \frac{t}{{t}_{r}}}\right) & \quad \mbox{for $
\left|{y}\right| \ge \frac{\lambda }{2}$.}\end{array}}\right.
\end{equation}
where $t_r = 5$, $\lambda = 30$, $L=200$ and $v_0 = 0.31$. This makes the axis of an emerging flux tube have a single $\Omega$-shape on the Sun.

Boundary conditions are the same as the ones used in Magara (2012). A periodic boundary is placed at $y=\pm 200$, while an open boundary is placed at $x=\pm 200$ and $z=190$. A fixed impermeable boundary is placed at $z=-10$. All these boundaries are accompanied by a wave damping zone, which avoids wave reflection at the boundaries. We enlarged a simulation domain compared to our previous work (Magara 2012) and terminated all the simulations before the outermost part of an emerging magnetic field reaches the top and side boundaries, so an emerging field expands in a simulation domain as if it expanded in an unbounded space. This makes it justified to apply the virial theorem given by Eq. (2) to our simulations.

To avoid the ambiguity of Equation (2) mentioned in the previous section, we make $\rho$, $P$, $v_z$, $B_x$, $B_y$ symmetric and $v_x$, $v_y$, $B_z$ anti-symmetric with respect to 180-degree rotation about the $z$-axis so that a magnetic field and an electric current are always unidirectional (Magara \& Longcope 2003). In this case,
\begin{equation}
{B}_{x}\left({x, y, z, t}\right) = {B}_{x}\left({- x, - y, z, t}\right),
\end{equation}
\begin{equation}
{B}_{y}\left({x, y, z, t}\right) = {B}_{y}\left({- x, - y, z, t}\right),
\end{equation}
and
\begin{equation}
{B}_{z}\left({x, y, z, t}\right) = - {B}_{z}\left({- x, - y, z, t}\right).
\end{equation}
According to Wheatland \& Metcalf (2006), the ambiguity of $E_{vr}$ is caused by non-zero values of the horizontal components of the net Lorentz force:
\begin{equation}
{F}_{x} = - \frac{1}{4\pi }\int_{z = {z}_{0}}^{}{B}_{x} {B}_{z} dx dy
\end{equation}
and
\begin{equation}
{F}_{y} = - \frac{1}{4\pi }\int_{z = {z}_{0}}^{}{B}_{y} {B}_{z} dx dy.
\end{equation}
Since Equations (11)-(13) make both $F_x$ and $F_y$ vanish, $E_{vr}$ gives a uniquely defined value in the present study.

\section{Results}

\subsection{Overview of emerging flux regions}

Firstly we present the overview of an emerging flux region in the EWT, WT, MT, ST and EST case. Figures 2a-2e show an emerging flux region obtained at a late phase in the EST (2a), ST (2b), MT (2c), WT (2d) and EWT (2e) case. In these figures, field lines are drawn in orange while isosurfaces of plasma $\beta$ = 1.0 and 0.1 are given in red and blue, respectively. Figure 3 shows the time variation of the emerged flux rate (Magara 2012), showing how much the net axial magnetic flux emerges into the solar atmosphere in each case. This figure suggests that flux emergence seems to be saturated at a late phase in all the cases, although it is difficult to reproduce complete saturation because it takes a significant amount of computational time toward a late phase. We therefore select the final time step of a simulation in each case to show a late-phase emerging flux region in Figure 2.

According to Murray et al. (2006), the smaller the twist parameter is, the less emerged magnetic flux is, which suggests that a low plasma $\beta$ region extends into the corona widely when the twist parameter is large. It is also found that a relatively high plasma $\beta$ region is locally formed in the corona when the twist parameter is large (see the EST case in Figure 2).

\subsection{Comparisons between $E_{vr}$ and $E_{m}$}

We then compare $E_{vr}$ and $E_{m}$ of an emerging flux region in these five cases. Figures 4a-c show how $E_{vr}$ (dashed line) and $E_{m}$ (solid line) change with time in the EST (black), ST (red), MT (green), WT (blue) and EWT (violet) cases, where $E_{vr}$ and $E_{m}$ are calculated by setting the bottom boundary at $z_0=0$ (4a), $z_0=1$ (4b) and $z_0=2$ (4c). From these figures, it is found that a difference between $E_{vr}$ and $E_{m}$ becomes small as $z_0$ increases, suggesting that a force-free approximation becomes reasonable when we use a magnetic field at a high atmospheric layer as a boundary condition. Another interesting result is that $E_{vr}$ tends to be negative during an early phase of flux emergence, which is discussed in Section 4.

We further investigate the force-free range of an emerging flux region in a quantitative way. Figure 5 shows how the difference between $E_{vr}$ and $E_{m}$ changes with $z_0$ in the EST case, obtained at $t=40$. Here we introduce the relative error between $E_{vr}$ and $E_{m}$ which is defined as
\begin{equation}
\Delta (z_0)= {|E_{m}(z_0) -E_{vr}(z_0)| \over E_{m}(z_0)}.
\label{eq:error}
\end{equation}
In Figure 5 $\Delta (z_0)$ first decreases sharply toward a local minimum and then it gradually increases with $z_0$. This suggests that an emerging magnetic field becomes close to a force-free state as it goes up from the photosphere while it is then deviated from a force-free state in an upper atmosphere where an emerging field is in a dynamic state, continuously expanding outward. The conversion of the magnetic energy into the kinetic energy that produces the expansion of envelop magnetic flux becomes significant toward a late phase during which the injection of the magnetic energy into the atmosphere becomes weak. This causes a peak and the succeeding decrease of the magnetic energy shown in Figure 2. 

The behavior of $\Delta (z_0)$ shown in Figure 5 is also found in the other cases once more than half of the net axial magnetic flux emerges into the photosphere. To determine the force-free range of an emerging flux region, we introduce a fitting function given by
\begin{equation}
f(z_0)=a e^{-h z_0}+b e^{k z_0},
\label{eq:fit}
\end{equation}
where $a$, $b$, $h$ and $k$ are constant. $f(z_0)$ takes a global minimum at $z_0=z_{min}$ given by
\begin{equation}
z_{min}={1 \over k+h} \ln \left({ha \over kb}\right).
\label{eq:min}
\end{equation}
A fitting result of $f(z_0)$ is shown by a red curve in Figure 5. Using $f(z_0)$, we define a force-free range as $z_l$ (lower limit) $\le z_0 \le z_u$ (upper limit) where $z_0$ satisfies $f(z_{0}) \le f_c \equiv e^{-1}$. We confirmed that $f(z_{min})$ is always below $f_c$ once more than half of the net axial magnetic flux emerges into the photosphere, which guarantees that we can determine a force-free range in all the cases. There is a tendency for $z_l$ to become close to $z_{min}$ as flux emergence proceeds.

Figures 6a and 6b show the time variations of $z_l$ (6a) and $z_u$ (6b) in the EST (black), ST (red), MT (green), WT (blue) and EWT (violet) case. $t=0$ corresponds to the time when more than half of the net axial magnetic flux emerges into the photosphere. $z_u$ tends to increase with time because the expansion of envelop magnetic flux makes a force-free range extend upward, while $z_l$ tends to decrease with time and then seems to be saturated because the expansion of magnetic flux toward a lower dense atmosphere is restricted. In general, as the twist parameter becomes large $z_l$ becomes small whereas $z_u$ becomes large, suggesting that an emerging flux region produced by a strongly twisted flux tube forms a wide force-free range.

Figure 7 shows the dependence of $z_l$ (solid line), $z_u$ (dashed line) and $z_{min}$ (dotted line) on the twist parameter. The values of $z_l$, $z_u$ and $z_{min}$ are chosen at the final time step of a simulation in each case. It is found that as the twist parameter becomes large $z_l$ decreases while $z_u$ increases. Both $z_l$ and $z_u$ seems to be saturated in the weakly twisted cases (WT and EWT case).

\section{Discussion}

According to Figure 7, an emerging magnetic field becomes force-free at an atmospheric layer whose height is the order of $H_{ph}$ (one $H_{ph}$ for the EST case and $4H_{ph}$ for the WT and EWT cases). A similar result is also reported in observational works where the surface integral of magnetic pressure is compared with the surface integral of each component of the Lorentz force (Low 1984) in active regions (Metcalf et al. 1995; Moon et al. 2002). The so-called magnetic canopy (Gabriel 1976) also has this height (Steiner 2001).  Figures 8a-c show how the ratio of the vertical component of the Lorentz force ($F_z$) to the magnetic pressure ($F_0$) changes with time as well as the location of the bottom boundary ($z_0$), both of which are given by
\begin{equation}
{F}_{z} = \frac{1}{8\pi }\int_{z = {z}_{0}}^{} \left({B}_{z}^2 - {B}_{x}^2 - B_y^2\right) dx dy,
\end{equation}
and
\begin{equation}
{F}_{0} = \frac{1}{8\pi }\int_{z = {z}_{0}}^{} \left({B}_{x}^2 + {B}_{y}^2 + B_z^2\right) dx dy.
\end{equation}
If the ratio is much smaller than 1, then a magnetic field is close to a force-free state (Metcalf et al. 1995). Figures 8a-c suggests that an emerging magnetic field tends to be force-free at a high atmospheric layer during a late phase. 

The twist of a flux tube makes an emerging magnetic field `rigid'; the magnetic tension force generated by a twisted field confines intense magnetic flux without the help of surrounding gas pressure, which means that a magnetic field is balanced by itself even in a high gas pressure region. This explains that an emerging magnetic field tends to be force-free at a low atmospheric layer with high gas pressure as the twist parameter increases.

Finally, let us discuss the behavior of $E_{vr}$ of an evolving magnetic structure formed via flux emergence. This is related to how much flux emergence proceeds, that is to say, the emerged flux rate plays a key role in determining the sign of $E_{vr}$. To show this, we use a two-dimensional model illustrated in Figure 9a. In this model
\begin{equation}
E_{vr} \propto \int_{z=z_0} Q dx,
\label{eq:ev2d}
\end{equation}
where $Q \equiv x B_x B_z$. The left and right panel in Figure 9a represents an early and late phase during which less than and more than half of the net axial magnetic flux emerges, respectively. A table given in lower part of each panel shows the signs of $x$, $B_z$, $B_x$ and $Q$ measured at the photosphere. These tables show that $Q$ mostly takes a negative value before half of the net axial magnetic flux emerges into the photosphere while positive $Q$ is found after more than half of the net axial magnetic flux emerges into the photosphere. Figure 9b shows a three-dimensional case obtained from one of the simulations (EST case). The photospheric distribution of $Q \equiv  (x B_x + y B_y) B_z$ is given at the left panel ($t=16$; before half of the net axial magnetic flux emerges) and right panel ($t=32$; after half of the net axial magnetic flux emerges). Here we plot several selected field lines in white. In Figure 9b the violet areas where $Q$ is negative are dominant during an early phase (left panel) while the red areas where $Q$ is positive become significant during a late phase (right panel). The transition from the early phase to the late phase shown in Figure 3 is related to the transition of $E_{vr}$ from decrease to increase found in Figure 4a. This result suggests that the virial theorem might be applied to a well-developed magnetic structure where flux emergence is almost saturated, although it may not be applied to a region where flux emergence actively proceeds.

%In summarizing this work, we have performed MHD simulations to produce emerging flux regions where we apply the virial theorem to investigate force-freeness in these regions. The lower limit of the force-free range corresponds to the order of the photospheric pressure scale height and it decreases as the twist of an emerging flux tube becomes strong. When the twist parameter decreases, the lower limit seems saturated. On the other hand, the upper limit of the range always increases, showing that the expansion of an emerging magnetic field continuously extends a force-free region upward.

\bigskip
The authors deeply appreciate useful comments by an anonymous referee. They also wish to thank the Kyung Hee University for general support of this work. KJH and TM give special thanks to Dr. Satoshi Inoue for his help. This work was financially supported by Basic Science Research Program (NRF-2013R1A1A2058705, PI: T. Magara) through the National Research Foundation of Korea (NRF) provided by the Ministry of Education, Science and Technology, as well as by the BK21 plus program through the NRF. Visualization in Figure 2 was done using VAPOR (Clyne et al. 2007).

\clearpage

\begin{figure}
\begin{center}
\includegraphics*[width=15cm]{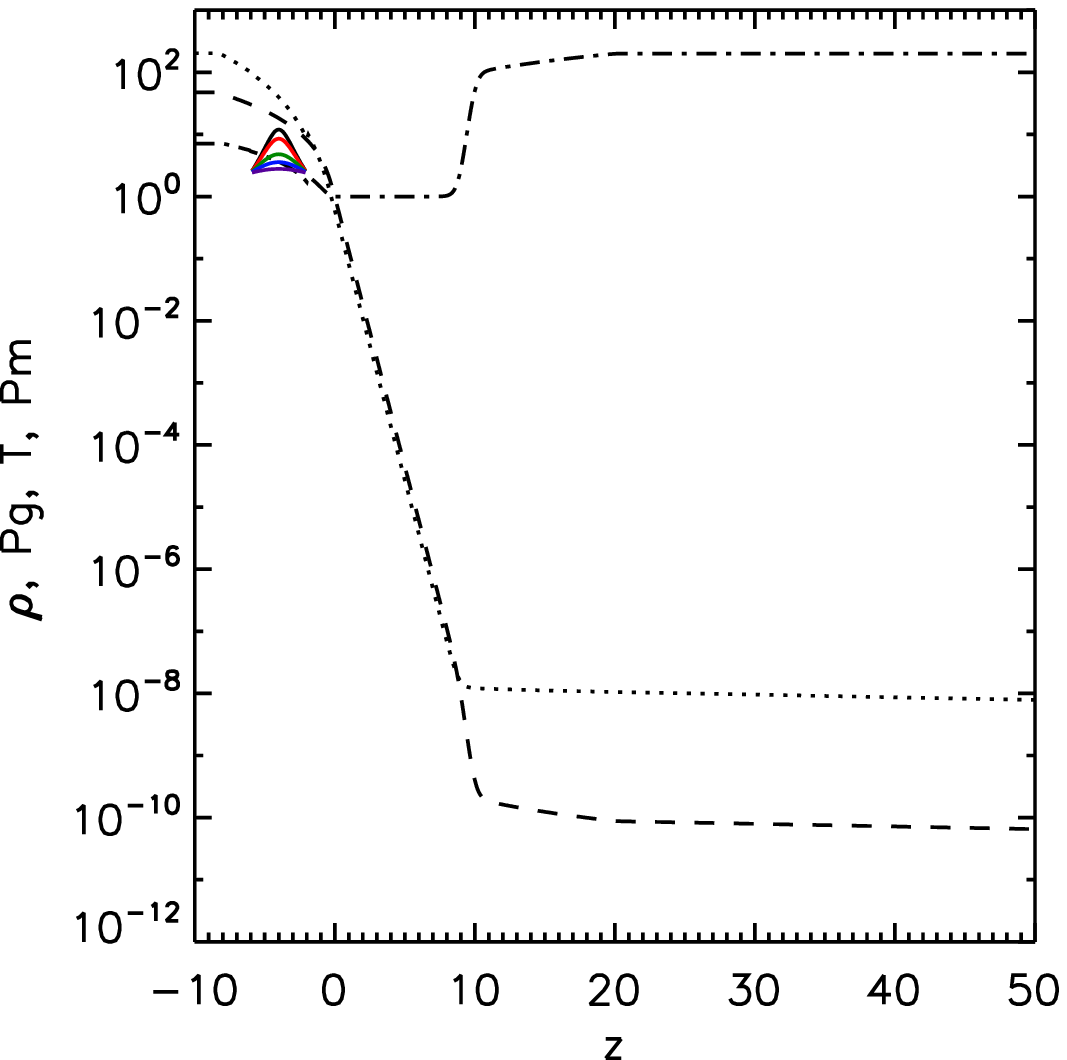}
\end{center}
\caption{{Initial distributions of physical quantities along $z$-axis are presented in logarithmic scale. These quantities are gas pressure (dotted line), density (dashed line), temperature (dot-dashed line) and magnetic pressure (solid line). The magnetic pressure of EST, ST, MT, WT and EWT case is drawn in black, red, green, blue and violet, respectively. The background atmosphere is common for all the cases.} 
\label{fig1}}
\end{figure}

\clearpage

\begin{figure}
\begin{center}
\includegraphics*[width=15cm]{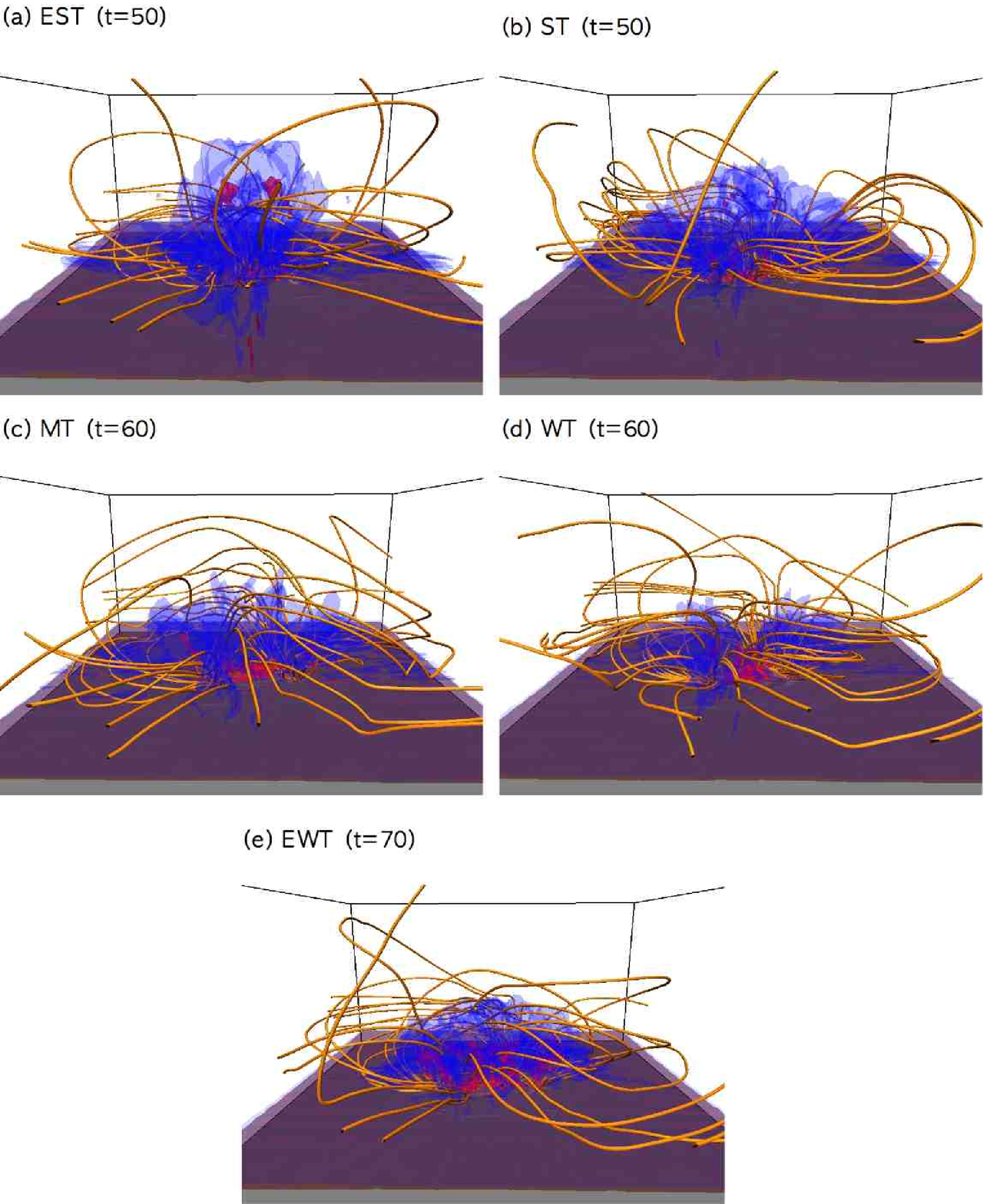}
\end{center}
\caption{{Emerging flux region is presented in EST (2a), ST (2b), MT (2c), WT (2d) and EWT (2e) case. Field lines are drawn in orange while an isosurface of plasma $\beta$ is given in red (plasma $\beta = 1$) and blue (plasma $\beta = 0.1$). The horizontal and vertical extents are $-80 \le x, y \le 80$ and $0 \le z \le 80$, respectively.} 
\label{fig2}}
\end{figure}

\clearpage

\begin{figure}
\begin{center}
\includegraphics*[width=15cm]{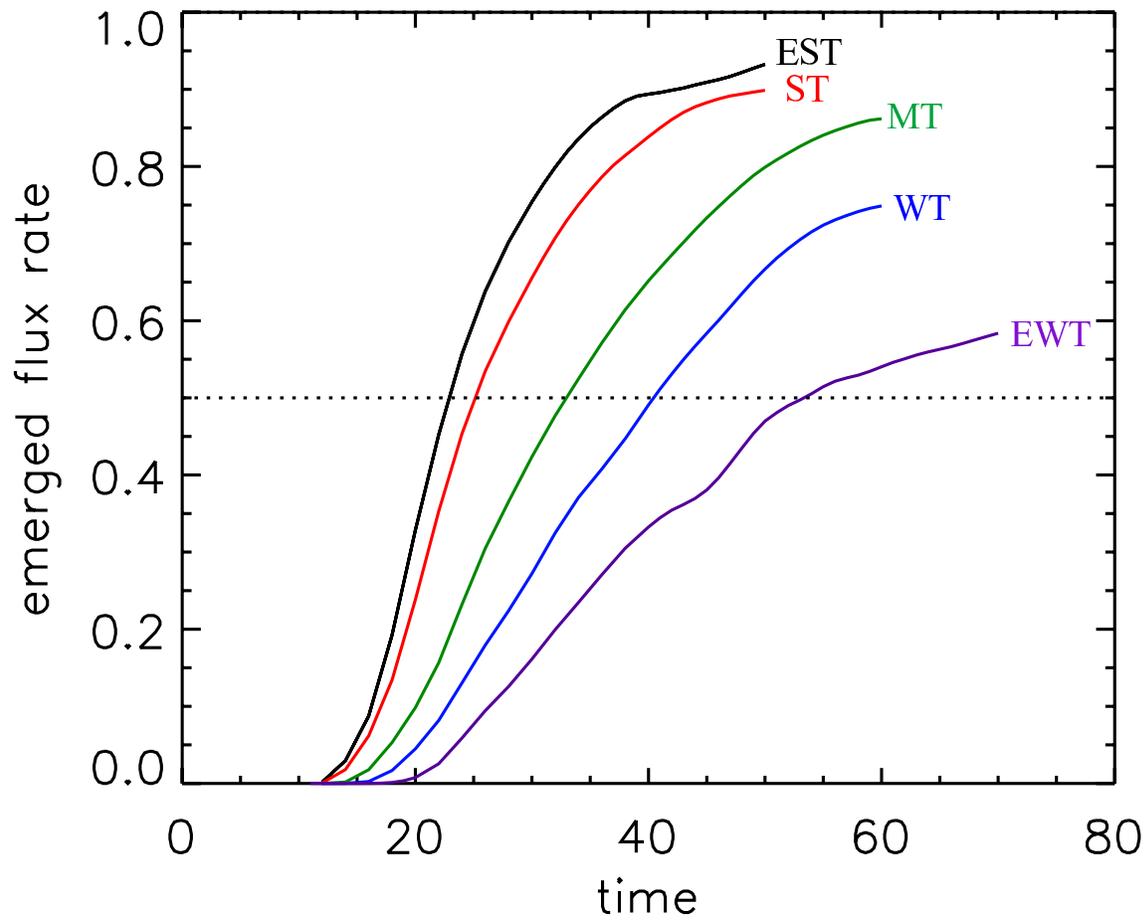}
\end{center}
\caption{{Time variations of emerged flux rate in EST (black), ST (red), MT (green), WT (blue) and EWT (violet) case.} 
\label{fig3}}
\end{figure}

\clearpage

\begin{figure}
\begin{center}
\includegraphics*[width=15cm]{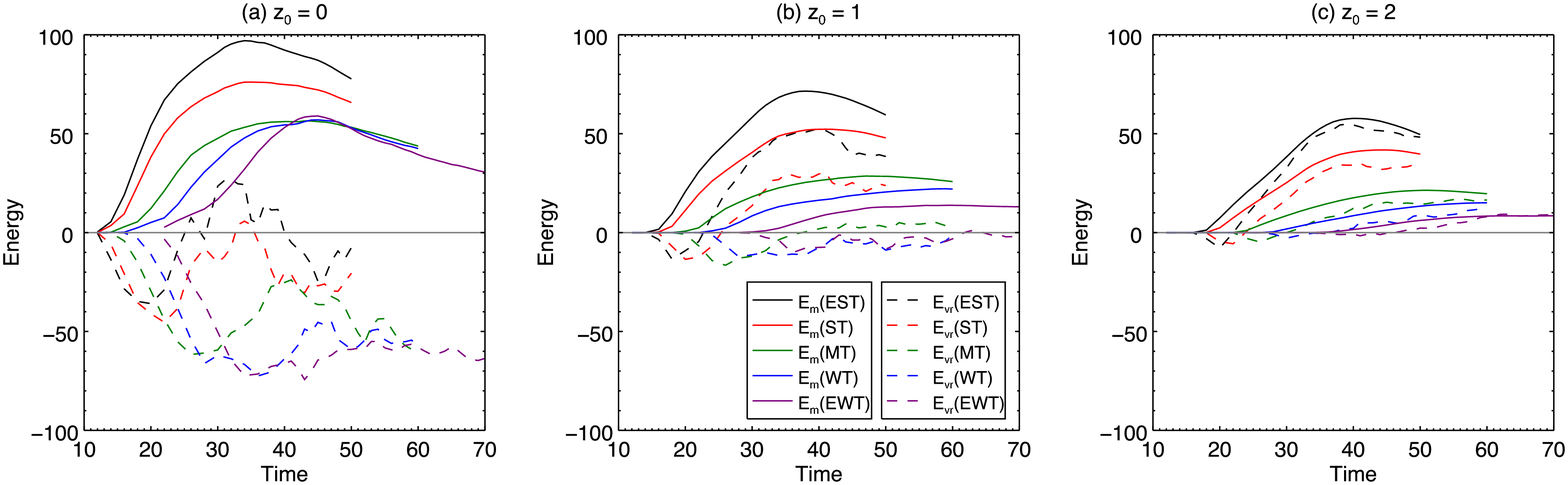}
\end{center}
\caption{{Time variations of $E_m$ (solid line) and $E_{vr}$ (dashed line) are presented in EST (black), ST (red), MT (green), WT (blue) and EWT (violet) case. The bottom boundary used to calculate them is placed at $z_0=0$ (4a), 1 (4b) and 2 (4c).} 
\label{fig4}}
\end{figure}

\clearpage

\begin{figure}
\begin{center}
\includegraphics*[width=15cm]{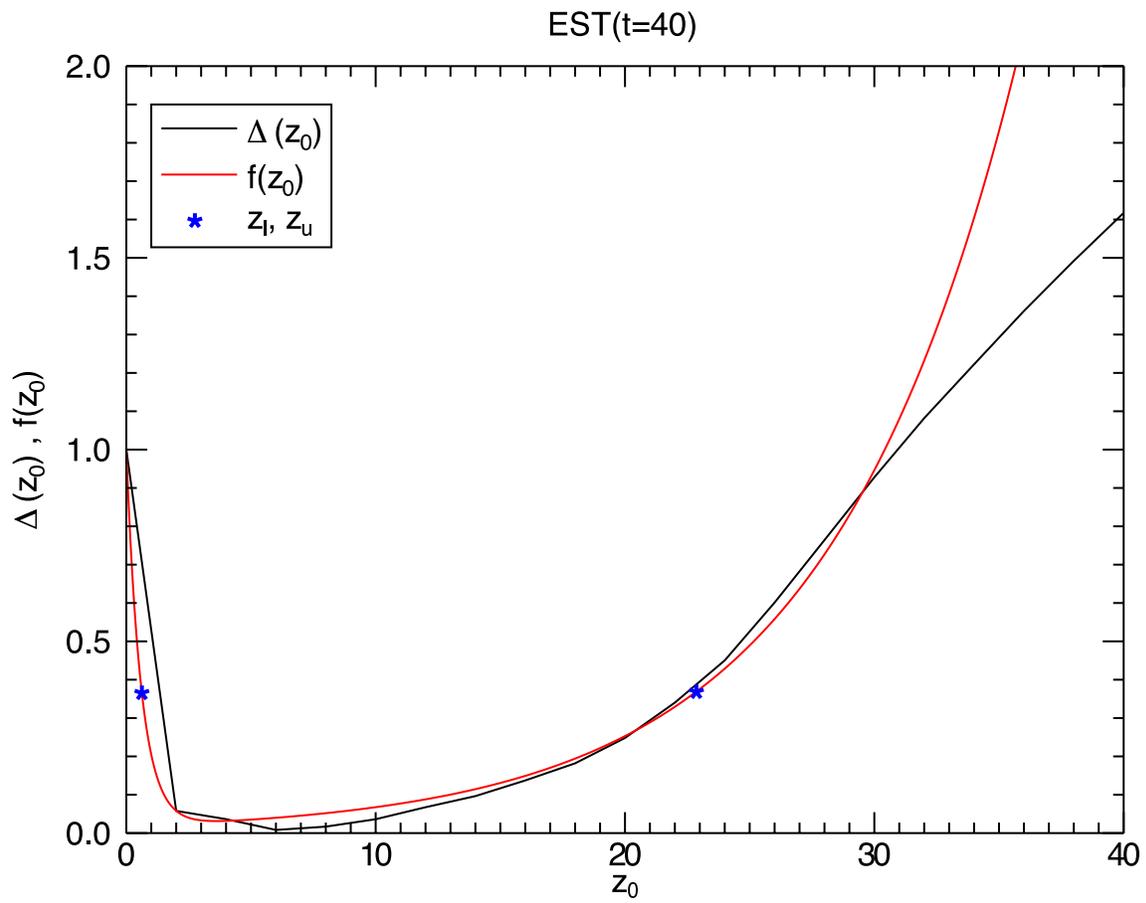}
\end{center}
\caption{{The solid line represents $\Delta (z_0)$ given by Equation (16) in EST case ($t=40$). A fitting curve given by Equation (17) is drawn in red. Two blue asterisks represent the locations of $z_0=z_l$ and $z_0=z_u$, respectively.} 
\label{fig5}}
\end{figure}

\clearpage

\begin{figure}
\begin{center}
\includegraphics*[width=15cm]{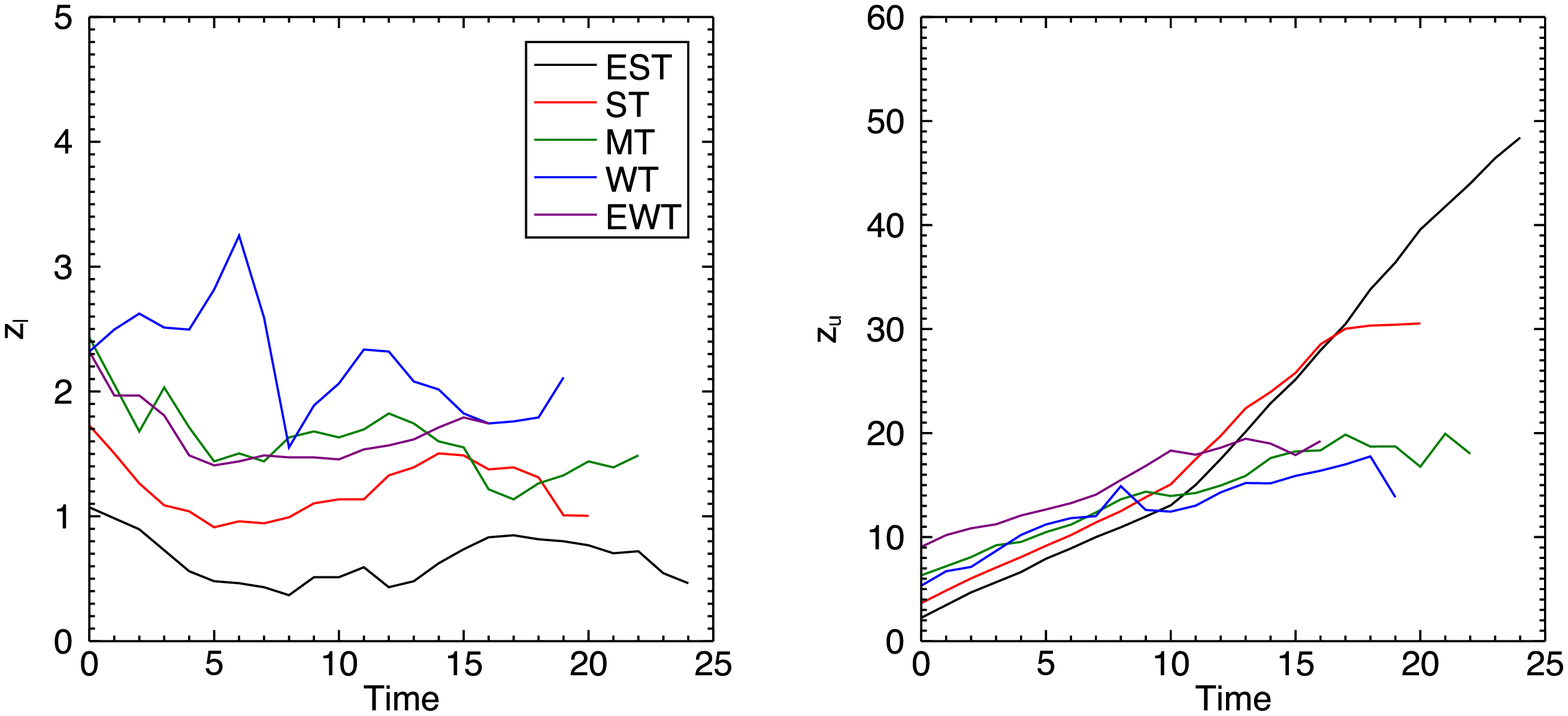}
\end{center}
\caption{{Time variations of $z_l$ (6a) and $z_u$ (6b) are presented in EST (black), ST (red), MT (green), WT (blue) and EWT (violet) case. $t=0$ corresponds to the time when more than half of the net axial magnetic flux emerges into the photosphere in each case.}
\label{fig6}}
\end{figure}

\clearpage

\begin{figure}
\begin{center}
\includegraphics*[width=15cm]{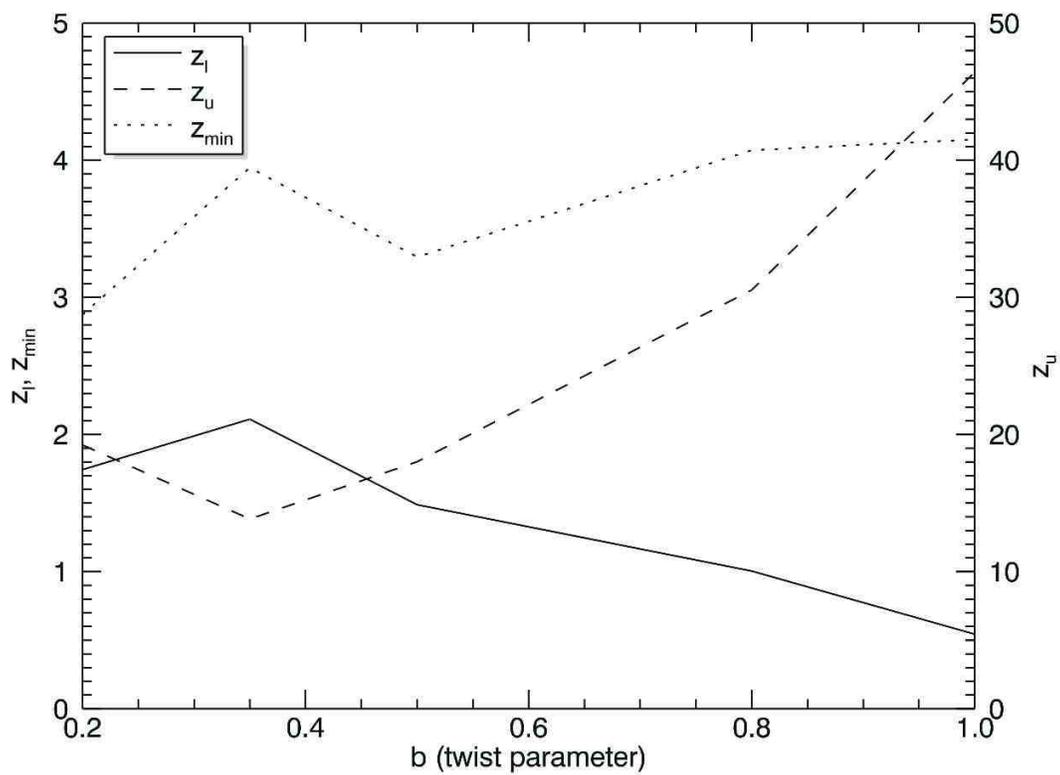}
\end{center}
\caption{{Dependence of $z_l$ (solid line, left vertical axis), $z_u$ (dashed line, right vertical axis) and $z_{min}$ (dotted line, left vertical axis) on twist parameter is presented in EWT ($b=0.2$), WT ($b=0.35$), MT ($b=0.5$), ST ($b=0.8$) and EST ($b=1$) case. The values of $z_l$, $z_u$ and $z_{min}$ are chosen at the final time step of a simulation in each case.} 
\label{fig7}}
\end{figure}

\clearpage

\begin{figure}
\begin{center}
\includegraphics*[width=15cm]{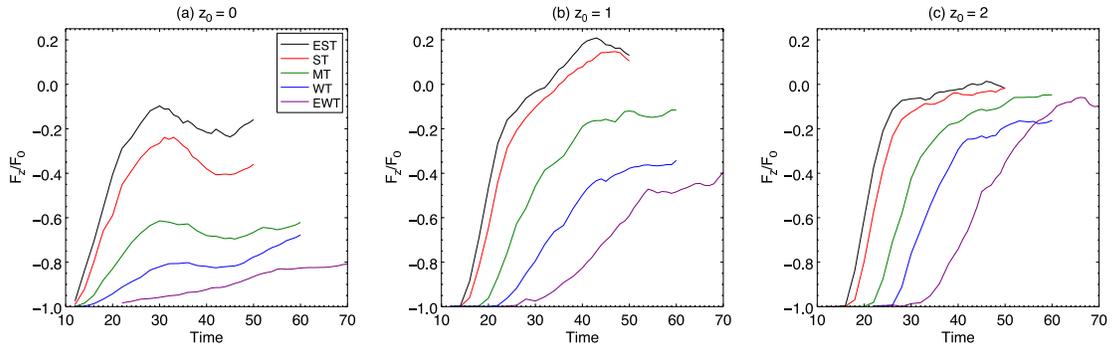}
\end{center}
\caption{{Time variation of $F_z/F_0$ given by Equations (19) and (20) is presented in EST (black), ST (red), MT (green), WT (blue) and EWT (violet) case. The bottom boundary used to calculate $F_z/F_0$ is placed at $z_0=0$ (8a), 1 (8b) and 2 (8c).} 
\label{fig8}}
\end{figure}

\clearpage

\begin{figure}
\begin{center}
\includegraphics*[width=12cm]{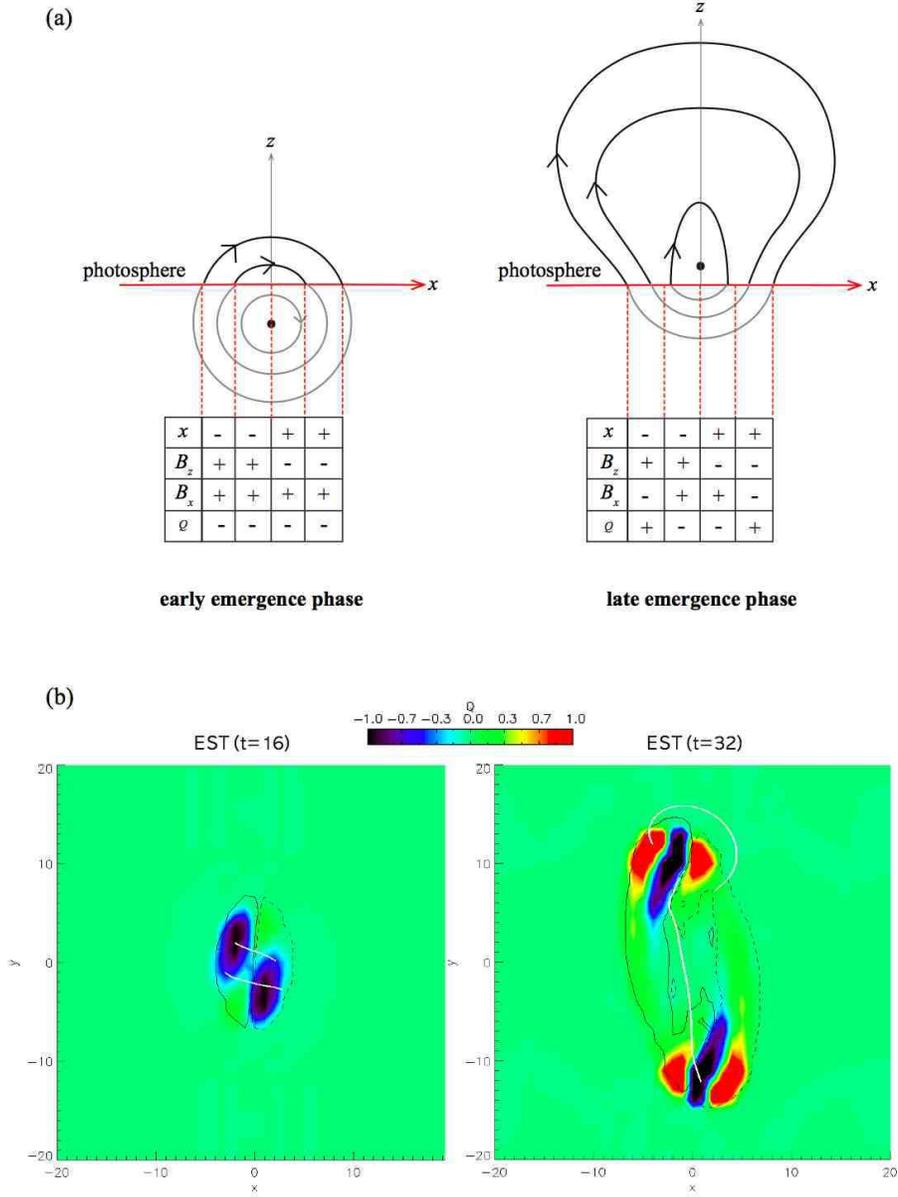}
\end{center}
\caption{{(a) Distribution of $Q \equiv x B_x B_z$ during an early and late phase is schematically illustrated using a two-dimensional model. The left panel shows an early phase when less than half of the net axial magnetic flux emerges into the photosphere while the right panel shows a late phase when more than half of it emerges into the photosphere. The signs of $x$, $B_z$, $B_x$ and $Q$ are given at lower part of each panel.
(b) Photospheric distribution of $Q \equiv  (x B_x + y B_y) B_z$ is presented by a color map at the left panel (early phase) and right panel (late phase) in EST case . Contours indicate $B_z =0.2$ (solid line) and $-0.2$ (dashed line). Several selected field lines are drawn in white.}
\label{fig9}}
\end{figure}

%
%\begin{figure}
%  \begin{center}
%\includegraphics*[width=15cm]{fig1.eps}
 % \end{center}
%\caption{Snapshots of an emerging flux tube in the strong twist (ST) case (left panels) and in the weak twist (WT) case (right panels).  The grey lines indicate magnetic field lines. At $t=0$, a magnetic flux tube in both cases is located below the photosphere at $z=0$ where the grey-scale map represents vertical magnetic flux. After emerging, the flux tube forms a helical structure of coronal magnetic field in the ST case while the coronal magnetic field has a diverging configuration in the WT case. \label{fig1}}
%\end{figure}

% Requires the booktabs if the memoir class is not being used

\clearpage

\begin{table}[htbp]
   \centering
   %\topcaption{Table captions are better up top} % requires the topcapt package
   \caption{Units of Physical Quantities}
   \begin{tabular}{@{} lcr @{}} % Column formatting, @{} suppresses leading/trailing space      
      \multicolumn{2}{c}{} \\
      \hline
      Physical Quantity & $Unit$\\
      \hline
	Length & 2$H_{ph}$ \footnotemark[a] \\
	Velocity & $c_{s_{ph}}$ \footnotemark[b] \\
	Time & 2$H_{ph}$ / $c_{s_{ph}}$ \\
	Gas Density & $\rho_{ph}$ \footnotemark[c] \\
	Gas Pressure & $\rho_{ph} c_{s_{ph}}^2$ \\
	Temperature & $T_{ph}$ \footnotemark[d] \\
	Magnetic Field & $(\rho_{ph} c_{s_{ph}}^2)^{1/2}$ \\
	\hline
   \end{tabular}
\end{table}
\footnotetext[a]{Photospheric gas pressure scale height.}
\footnotetext[b]{Photospheric adiabatic sound speed.}
\footnotetext[c]{Photospheric gas density.}
\footnotetext[d]{Photospheric temperature.}

\clearpage

\begin{table}
\centering
\caption{Parameters used in the simulations \label{tbl-1}}
\begin{tabular}{lllllll}
%%{p{3cm} p{2cm} p{2cm} p{2cm} p{2cm} p{2cm} p{2cm}} 
   \hline
Case & $r_f$ \footnotemark[a] & $\lambda$ & $l$ \footnotemark[b] & $b$ & $B_0$ & $\Phi_0$ \footnotemark[c] \\
\hline
EST   & 2     & 30   & 6.3   & 1     & 17  & 88 \\
ST    & 2     & 30   & 7.9   & 0.8   & 15  & 91 \\
MT    & 2     & 30   & 13   & 0.5    & 11 & 96 \\
WT    & 2     & 30   & 18   & 0.35   & 9.5 & 97 \\
EWT   & 2     & 30   & 31   & 0.2    & 8.4 & 98 \\
\hline
\end{tabular}
\end{table}
\footnotetext[a]{Radius of a flux tube.}
\footnotetext[b]{$l \equiv 2 \pi /b$, representing the axial distance of a field line that makes one helical turn around flux-tube axis.}
\footnotetext[c]{Net axial magnetic flux.}

%\begin{table}[htbp]
%   \centering
%   %\topcaption{Table captions are better up top} % requires the topcapt package
%   \begin{tabular}{@{} lcr @{}} % Column formatting, @{} suppresses leading/trailing space      
%      \multicolumn{2}{c}{} \\
%      \hline
%      Physical Quantity & $Unit$\\
%      \hline
%	Length & 2 $\Lambda_{ph}$ \\
%	Velocity & $c_{s_{ph}}$\\
%	Time & 2 $\Lambda_{ph}$ / $c_{s_{ph}}$ \\
%	Gas Density & $\rho_{ph}$\\
%	Gas Pressure & $\rho_{ph} c_{s_{ph}}^2$ \\
%	Temperature & $T_{ph}$ \\
%	Magnetic Field & $(\rho_{ph} c_{s_{ph}}^2)^{1/2}$ \\
%	\hline
%   \end{tabular}
%
%   \caption{Units of Physical Quantities}
%   \label{table:units}
%\end{table}

\end{document}